\documentclass[prd,aps,nofootinbib,superscriptaddress,floatfix,preprintnumbers,twocolumn]{revtex4}
\usepackage{hyperref,amssymb,amsmath,mathrsfs,amsthm,graphicx,xcolor,slashed}

\begin{document}
\preprint{MSUHEP-21-033}

\title{First Glimpse into the Kaon Gluon Parton Distribution Using Lattice QCD}

\author{Alejandro Salas-Chavira}
\email{salasale@msu.edu}
\affiliation{Department of Physics and Astronomy, Michigan State University, East Lansing, MI 48824}

\author{Zhouyou Fan}
\email{fanzhouy@msu.edu}
\affiliation{Department of Physics and Astronomy, Michigan State University, East Lansing, MI 48824}

\author{Huey-Wen Lin}
\email{hwlin@pa.msu.edu}
\affiliation{Department of Physics and Astronomy, Michigan State University, East Lansing, MI 48824}
\affiliation{Department of Computational Mathematics,
  Science and Engineering, Michigan State University, East Lansing, MI 48824}

\begin{abstract}
In this work, we present the first results on the gluon parton distribution of the kaon from lattice quantum chromodynamics.
We carry out the lattice calculation at pion mass around 310~MeV and two lattice spacings, 0.15 and 0.12~fm, using $2+1+1$-flavor HISQ ensembles generated by MILC Collaboration.
The kaon correlators are calculated using clover fermions and momentum-smearing sources with maximum boost momentum around 2~GeV and high statistics (up to 324,000 measurements).
We study the dependence of the resulting reduced Ioffe-time pseudo-distributions at multiple boost momenta and lattice spacings.
We then extract the kaon gluon distribution function in the $\overline{\text{MS}}$ scheme at $\mu = 2$~GeV, neglecting the mixing between the gluon and singlet-quark sectors.
Our results at the smaller lattice spacing are consistent with phenomenological determinations.
\end{abstract}

\maketitle

\section{Introduction}

The study of pseudoscalar meson structure could shed light into the longstanding quantum-chromodynamics (QCD) mystery of the origin of mass.
In the Standard Model, the Higgs boson provides most of the masses of all noncomposite fermions.
However, the masses of the valence quarks of mesons contribute only a small fraction of the total meson mass.
The remainder is thought to arise from the dynamics of the strong interaction.
Studying kaon and pion structure can greatly aid our understanding of the emergence of hadronic mass.
One commonly studied type of structure is the parton distribution function (PDF), which gives the probability to find quarks and gluons inside a hadron as a function of their momentum fraction $x$ of the mother hadron.
Since the gluon field is responsible for almost all of the pion's mass, studying the gluon PDFs of mesons would be very valuable in understanding the structure of mesons.
There have been a number of attempts to extract pion gluon PDFs~\cite{Barry:2018ort,Cao:2021aci,Novikov:2020snp,Chang:2020rdy}, but not much progress has been made on the kaon PDFs.
The future experimental studies, for example, U.S.-based Electron-Ion Collider (EIC)~\cite{Accardi:2012qut,AbdulKhalek:2021gbh}, the Electron-Ion Collider in China (EicC)~\cite{Anderle:2021wcy},
the Drell-Yan and $J/\psi$-production experiments  COMPASS++/AMBER~\cite{Denisov:2018unj} in Europe  will aim at improving our knowledge of both the pion and kaon gluon and quark PDFs. 
We refer readers to recent reviews~\cite{Aguilar:2019teb,Roberts:2021nhw,Arrington:2021biu} for more details on the current and future prospects for studies of pion and kaon structure.

Lattice QCD provides a method for determining Standard-Model inputs to improve our knowledge of nonperturbative meson gluon structure.
However, due to the notorious noise-to-signal issue associated with gluon observables, there have been only a few efforts to determine the first moment of pion gluon PDF~\cite{Meyer:2007tm,Shanahan:2018pib} with 450~MeV as the lightest pion mass calculated.
No kaon moments have yet been calculated.
There is little chance that we will be able to use lattice calculations of increasingly higher moments to reconstruct the $x$-dependence of the meson PDFs. 
Recently, direct lattice calculations of the gluon PDF of hadrons have been proposed: pseudo-PDF~\cite{Balitsky:2019krf} and quasi-PDF~\cite{Zhang:2018diq,Wang:2019tgg} approaches, following the proposal of Large-Momentum Effective Theory (LaMET)~\cite{Ji:2013dva,Ji:2014gla,Ji:2017rah}.
Ongoing efforts using other methods, such as hadronic tensor~\cite{Liu:1993cv}, the operator product expansion~\cite{Detmold:2005gg}, lattice cross sections~\cite{Ma:2014jla,Ma:2017pxb} and the pseudo-PDF approach~\cite{Radyushkin:2017cyf,Orginos:2017kos}, have also had some success in extracting $x$-dependent hadron structure.
Most such PDF calculations, however, are done using the popular quasi-PDF and pseudo-PDF techniques and mostly limited to isovector quark distributions in the nucleon and the valence-quark distribution in the pion and kaons~\cite{Lin:2013yra,Lin:2014zya,Chen:2016utp,Lin:2017ani,Alexandrou:2015rja,Alexandrou:2016jqi,Alexandrou:2017huk,Chen:2017mzz,Alexandrou:2018pbm,Chen:2018xof,Chen:2018fwa,Alexandrou:2018eet,Lin:2018qky,Fan:2018dxu,Liu:2018hxv,Wang:2019tgg,Lin:2019ocg,Chen:2019lcm,Lin:2020reh,Chai:2020nxw,Bhattacharya:2020cen,Lin:2020ssv,Zhang:2020dkn,Li:2020xml,Fan:2020nzz,Gao:2020ito,Lin:2020fsj,Lin:2021brq,Zhang:2020rsx,Alexandrou:2020qtt,Alexandrou:2020zbe,Lin:2020rxa,Gao:2021hxl,Lin:2020rut,Orginos:2017kos,Karpie:2017bzm,Karpie:2018zaz,Karpie:2019eiq,Joo:2019jct,Joo:2019bzr,Radyushkin:2018cvn,Zhang:2018ggy,Izubuchi:2018srq,Joo:2020spy,Bhat:2020ktg,Fan:2020cpa,Sufian:2020wcv,Karthik:2021qwz,Chen:2018fwa,Sufian:2019bol,Izubuchi:2019lyk,Joo:2019bzr,Sufian:2020vzb,Shugert:2020tgq,Gao:2020ito}.

The first exploratory study applying the quasi-PDF approach to gluon PDFs~\cite{Fan:2018dxu} used overlap valence fermions on gauge ensembles with 2+1 flavors of domain-wall fermion and spacetime volume $24^3\times 64$, $a=0.1105(3)$~fm, and $M_\pi^\text{sea}=330$~MeV.
The gluon operators were calculated for all spacetime lattice sites and high statistics: 207,872 measurements were taken of the two-point functions with valence quarks at the light sea and strange masses (corresponding to pion masses 340 and 678~MeV, respectively).
Both pion and nucleon coordinate-space gluon quasi-PDF matrix-element ratios were compared with global fits after Fourier transformation;
the lattice results were consistent with global fits within large uncertainties. 
Further progress was made using the pseudo-PDF method.
Fan et~al. studied the nucleon gluon PDF using clover fermions on $N_f=2+1+1$ MILC lattices, with a single lattice spacing 0.12~fm and two valence pion masses: 310 and 690~MeV~\cite{Fan:2020cpa}. 
This was the first $x$-dependent gluon PDF calculated using lattice QCD. 
Khan et~al. used distillation nucleon interpolating operators and extracted the nucleon gluon PDF on a 2+1-flavor 358-MeV pion mass ensemble at lattice spacing 0.094~fm~\cite{HadStruc:2021wmh}.
Both works neglected the mixing of the gluon operator with the quark-singlet sector.
Since then, Fan has continued to pursue gluon PDF calculations of the nucleon and pion using $N_f=2+1+1$ MILC lattices with multiple lattice spacings and pion masses. 
Reference~\cite{Fan:2021bcr} reported the first lattice-QCD pion $x$-dependent PDFs at two lattice spacings: 0.15 and 0.12~fm and $M_\pi \approx 220$, 310 and 690~MeV);
the work used the global-fit PDF quark contribution to estimate the effect of the quark mixings in the gluon PDF determination.
However, kaon PDF calculations remain limited to the valence-quark distribution~\cite{Lin:2020ssv}.

In this work, we present the first lattice-QCD calculation of the $x$-dependent kaon gluon parton distribution using the pseudo-PDF method.
We study the kaon PDFs using two lattice spacings, 0.12 and 0.15~fm, and  one pion mass of 310~MeV.
The rest of the paper is organized as follows.
In Sec.~\ref{sec:lattice-details}, we discuss the operator used in this calculation, the numerical setup of lattice simulation, and how the ground-state kaon gluon matrix elements are obtained on the lattice.
In Sec.~\ref{sec:results}, we discuss our strategy for extracting the kaon gluon PDF from our lattice calculations.
The systematic uncertainties introduced by different steps are studied, and the lattice-spacing and pion-mass dependence are investigated.

\section{Lattice Matrix Elements}\label{sec:lattice-details}

The lattice gluon matrix elements of the kaon's ground-state  $|0 (P_z) \rangle$  are calculated for several boost momenta $P_z$ and Wilson-line displacement lengths $z$ using the unpolarized gluon operator first introduced in Ref.~\cite{Balitsky:2019krf},
\begin{equation}\label{eq:gluon_operator}
 {\cal O}(z)\equiv\sum_{i\neq z,t}{\cal O}(F^{ti},F^{ti};z)-\sum_{i,j\neq z,t}{\cal O}(F^{ij},F^{ij};z),
\end{equation}
where ${\cal O}(F^{\mu\nu}, F^{\alpha\beta};z) = F^\mu_\nu(z)U(z,0)F^\alpha_\beta(0)$, the Wilson link length is denoted by $z$, and the field strength $F_{\mu\nu}$ is given by
\begin{equation}\label{field_strength}
 F_{\mu\nu} = \frac{i}{8a^2g_0}\left(\mathcal{P}_{[\mu,\nu]}+\mathcal{P}_{[\nu,-\mu]}+\mathcal{P}_{[-\mu,-\nu]}+\mathcal{P}_{[-\nu,\mu]}\right),
\end{equation}
where $a$ is the lattice spacing, $g_0$ is the strong coupling constant, with the plaquette $\mathcal{P_{\mu,\nu}}=U_\mu(x)U_\nu(x+a\hat{\mu})U^\dag_\mu(x+a\hat{\nu})U^\dag_\nu(x)$ and $\mathcal{P_{[\mu,\nu]}}=\mathcal{P}_{\mu,\nu}-\mathcal{P}_{\nu,\mu}$.
The operator $\sum_{i\neq z,t}{\cal O}(F^{ti},F^{zi};z)$ corresponds to the same matching kernel as in Ref.~\cite{Balitsky:2019krf}.
However, this operator is not employed, since it vanishes at $P_z=0$ due to kinematic reasons, which increases the difficulty of obtaining the distributions.
Having calculated the aforementioned matrix elements, we investigate their dependence on Ioffe time, $\nu=zP_z$.

On the lattice, we employ clover valence fermions on three ensembles with $N_f=2+1+1$ highly improved staggered quarks (HISQ)~\cite{Follana:2006rc}, generated by the MILC Collaboration~\cite{MILC:2010pul,Bazavov:2012xda} at two lattice spacings ($a\approx 0.12$ and 0.15~fm) and pion mass of 310~MeV.
The clover quark masses were adjusted to reproduce the lightest strange and light sea pseudoscalar meson utilized by the PNDME collaboration~\cite{Rajan:2017lxk,Bhattacharya:2015wna,Bhattacharya:2015esa,Bhattacharya:2013ehc}.
We apply five steps of HYP smearing~\cite{Hasenfratz:2001hp} for the gluon loops to reduce the statistical noise, as elucidated in Ref.~\cite{Fan:2018dxu}.
In order to reach higher meson boost momenta, we use Gaussian momentum smearing~\cite{Bali:2016lva} for the quark fields.
Table~\ref{table-data} shows the lattice spacing $a$, valence pion mass $M_\pi^\text{val}$ and kaon mass $M_K^\text{val}$, lattice size $L^3\times T$, number of configurations $N_\text{cfg}$, and number of total two-point correlator measurements $N_\text{meas}^\text{2pt}$ 
for the three ensembles.

\begin{table}[!htbp]
\centering
\begin{tabular}{|c|c|c|}
\hline
  ensemble &  a12m310 & a15m310 \\
\hline
  $a$ (fm)   & $0.1207(11)$  & $0.1510(20)$ \\
\hline
  $M_\pi^\text{val}$ (MeV)  & $311.1(6)$ & $319.1(31)$\\
\hline
  $M_{k}^\text{val}$ (MeV)  & 528.0(5) &  531.7(23)  \\
\hline
  $L^3\times T$   & $24^3\times 64$ & $16^3\times 48$ \\
\hline
  $P_z$ (GeV)   &  $[0,2.14]$ &  $[0,2.05]$ \\
\hline
  $N_\text{cfg}$  & 1013 & 900 \\
\hline
  $N_\text{meas}^\text{2pt}$   & 324,160 & 21,600 \\
\hline
\end{tabular}
\caption{
Lattice spacing $a$, valence pion mass $M_\pi^\text{val}$ and kaon mass $M_{k}^\text{val}$, lattice size $L^3\times T$, number of configurations $N_\text{cfg}$, number of total two-point correlator measurements $N_\text{meas}^\text{2pt}$, and separation times $t_\text{sep}$ utilized in three-point correlator fits of $N_f=2+1+1$ clover valence fermions on HISQ ensembles generated by MILC Collaboration and analyzed in this study.
}\label{table-data}
\end{table}

For a meson $\Phi$, the two-point correlator is given by
\begin{align}
C_\Phi^\text{2pt}(P_z;t)&=
 \int d^3y\, e^{-i y\cdot P_z} \langle \chi_\Phi(\vec{y},t)|\chi_\Phi(\vec{0},0)\rangle \nonumber \\
  &= |A_{\Phi,0}|^2 e^{-E_{\Phi,0}t} + |A_{\Phi,1}|^2 e^{-E_{\Phi,1}t} + ...,
\label{eq:2pt_fit_formula}
\end{align}
where $P_z$ is the meson momentum in the $z$-direction, $\chi_\Phi=\bar{q}_1\gamma_5 q_2$ is the pseudoscalar-meson interpolation operator, and $t$ is the Euclidean time.
$|A_{\Phi,i}|^2$ and $E_{\Phi,i}$ denote the amplitude and energy for the ground state ($i=0$) and the first excited state ($i=1$), respectively.
We generate the three-point gluon correlators by multiplying the kaon two-point correlators with the gluon operators.
The matrix elements of the gluon operators are then extracted by fitting the three-point correlators to the first terms of the energy-eigenstate expansion,
\begin{align}\label{eq:3ptC}
&C_\Phi^\text{3pt}(z,P_z;t_\text{sep},t) \nonumber \\
&=\int d^3y\, e^{-iy\cdot P_z}\langle \chi_\Phi(\vec y,t_\text{sep})|{\cal O}(z,t)|\chi_\Phi(\vec 0,0)\rangle \nonumber \\
&= |A_{\Phi,0}|^2\langle 0|{\cal O}|0\rangle e^{-E_{\Phi,0}t_\text{sep}} \nonumber \\
&+ |A_{\Phi,0}||A_{\Phi,1}|\langle 0|{\cal O}|1\rangle e^{-E_{\Phi,1}(t_\text{sep}-t)}e^{-E_{\Phi,0}t} \nonumber \\
&+ |A_{\Phi,0}||A_{\Phi,1}|\langle 1|{\cal O}|0\rangle e^{-E_{\Phi,0}(t_\text{sep}-t)}e^{-E_{\Phi,1}t} \nonumber \\
&+ |A_{\Phi,1}|^2\langle 1|{\cal O}|1\rangle e^{-E_{\Phi,1}t_\text{sep}}+ ...,
\end{align}
where $t_{\text{sep}}$ is the source-sink time separation, and $t$ is the gluon-operator insertion time.
The amplitudes $A_{\Phi,0}$, $A_{\Phi,1}$, and energies $E_{\Phi,0}$, $E_{\Phi,1}$ are extracted from two-state fits to the two-point correlators.
$\langle 0|{\cal O}|0\rangle$, $\langle 0|{\cal O}|1\rangle$ ($\langle 1|{\cal O}|0\rangle$), and $\langle 1|{\cal O}|1\rangle$ denote the ground-state matrix element, the ground-excited matrix element, and the excited-state matrix element, respectively.

\begin{figure*}[!tbp]
  \centering
\includegraphics[width=0.35\textwidth]{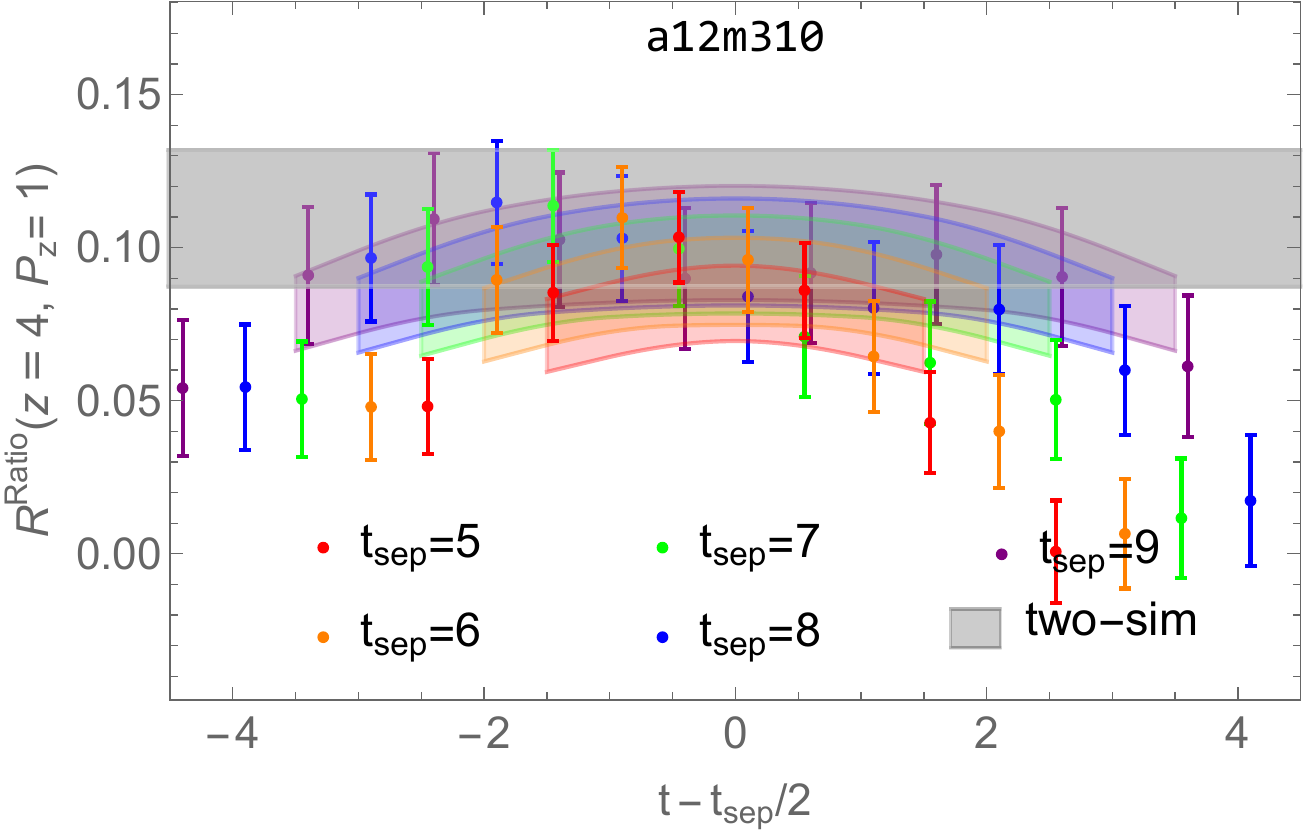}
\includegraphics[width=0.2\textwidth]{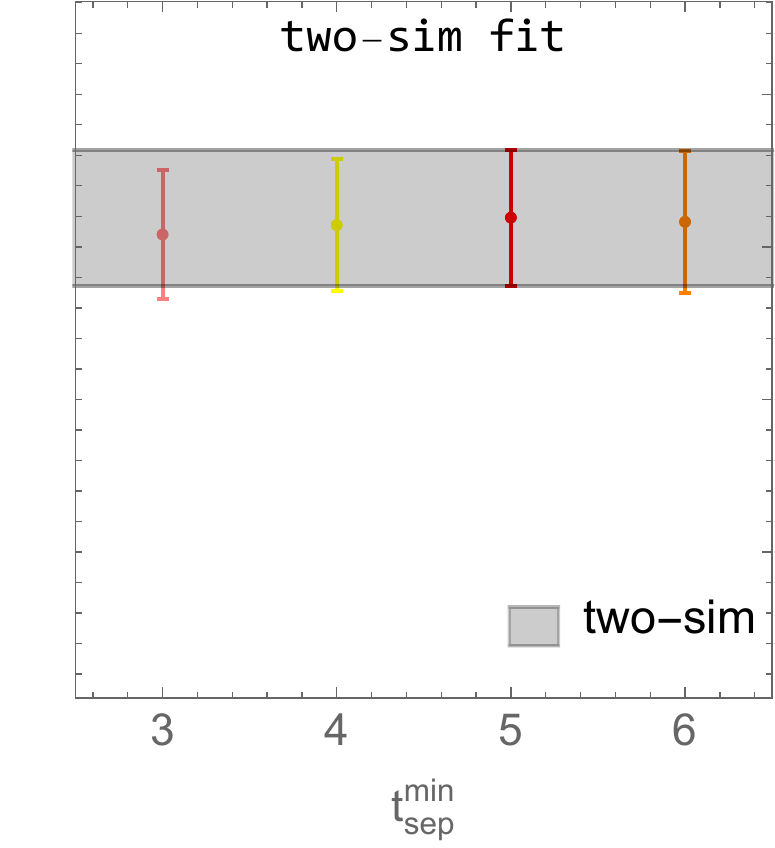}
\includegraphics[width=0.2\textwidth]{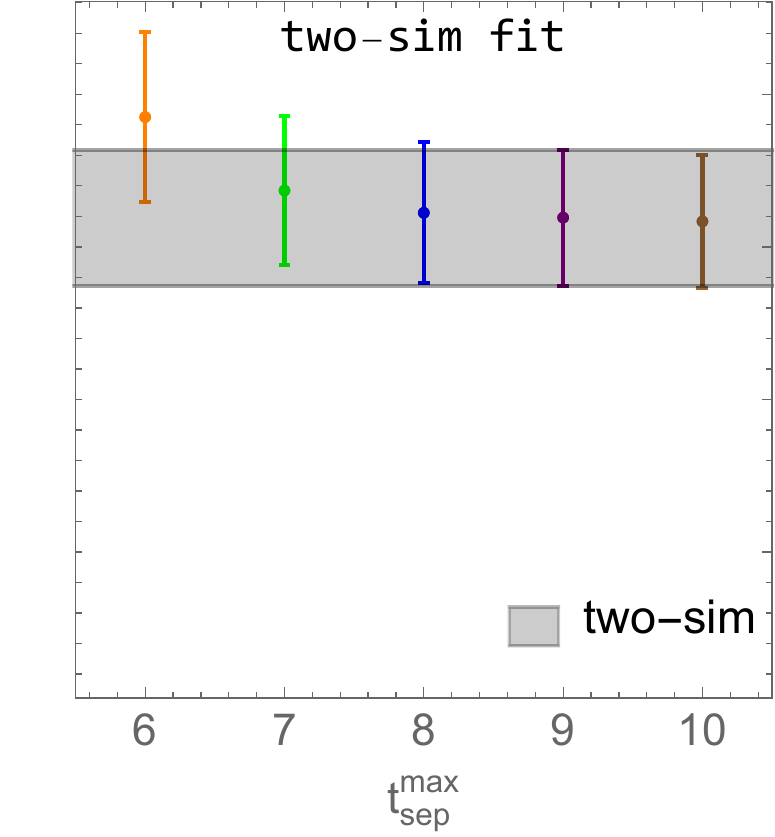}
\includegraphics[width=0.35\textwidth]{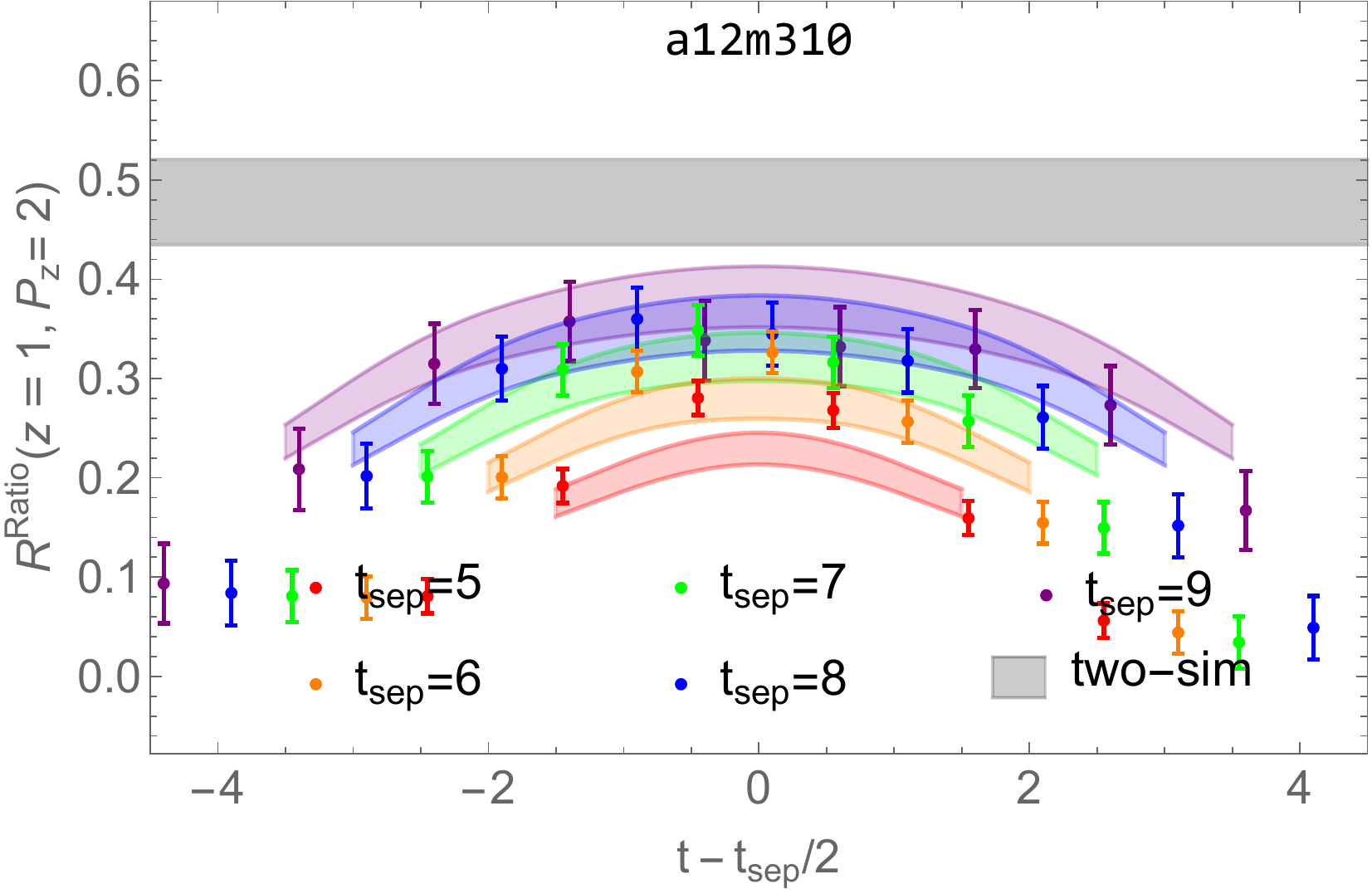}
\includegraphics[width=0.2\textwidth]{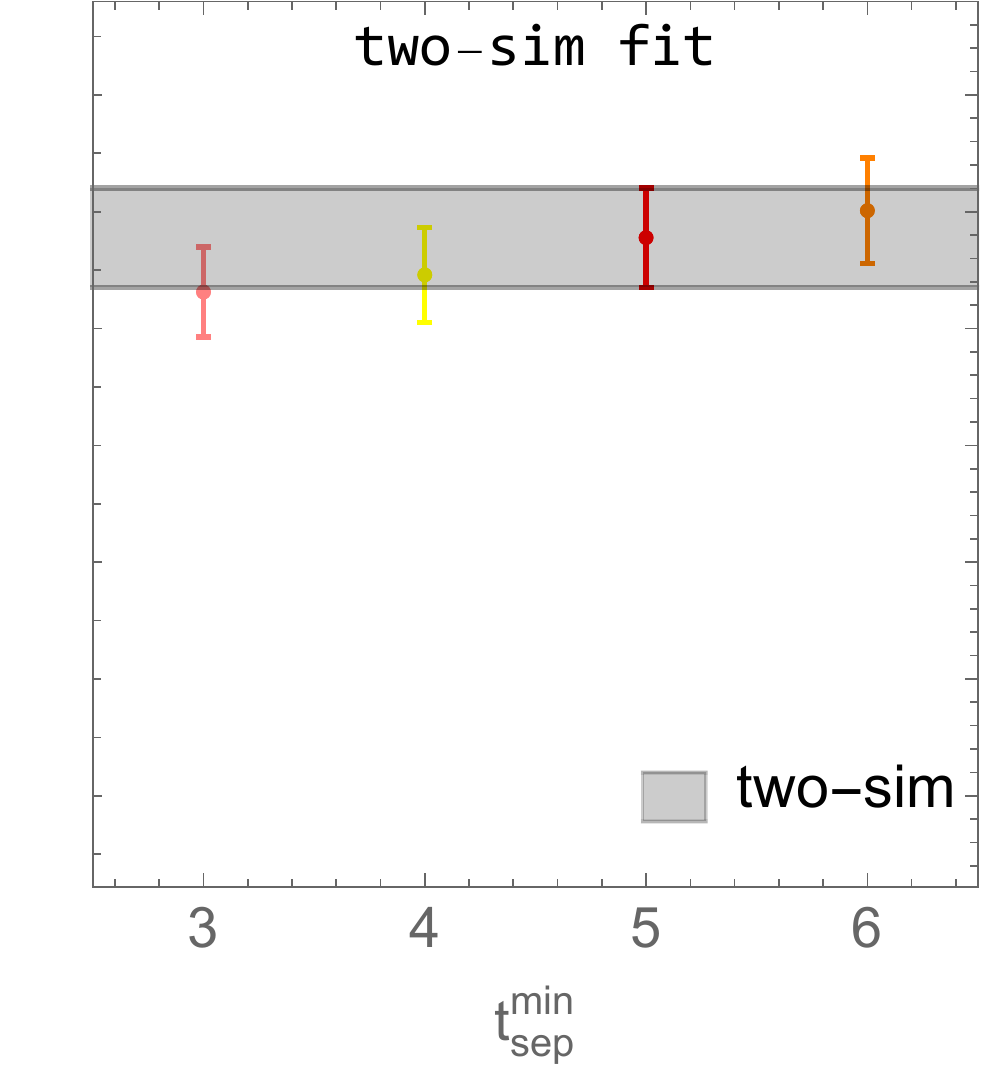}
\includegraphics[width=0.2\textwidth]{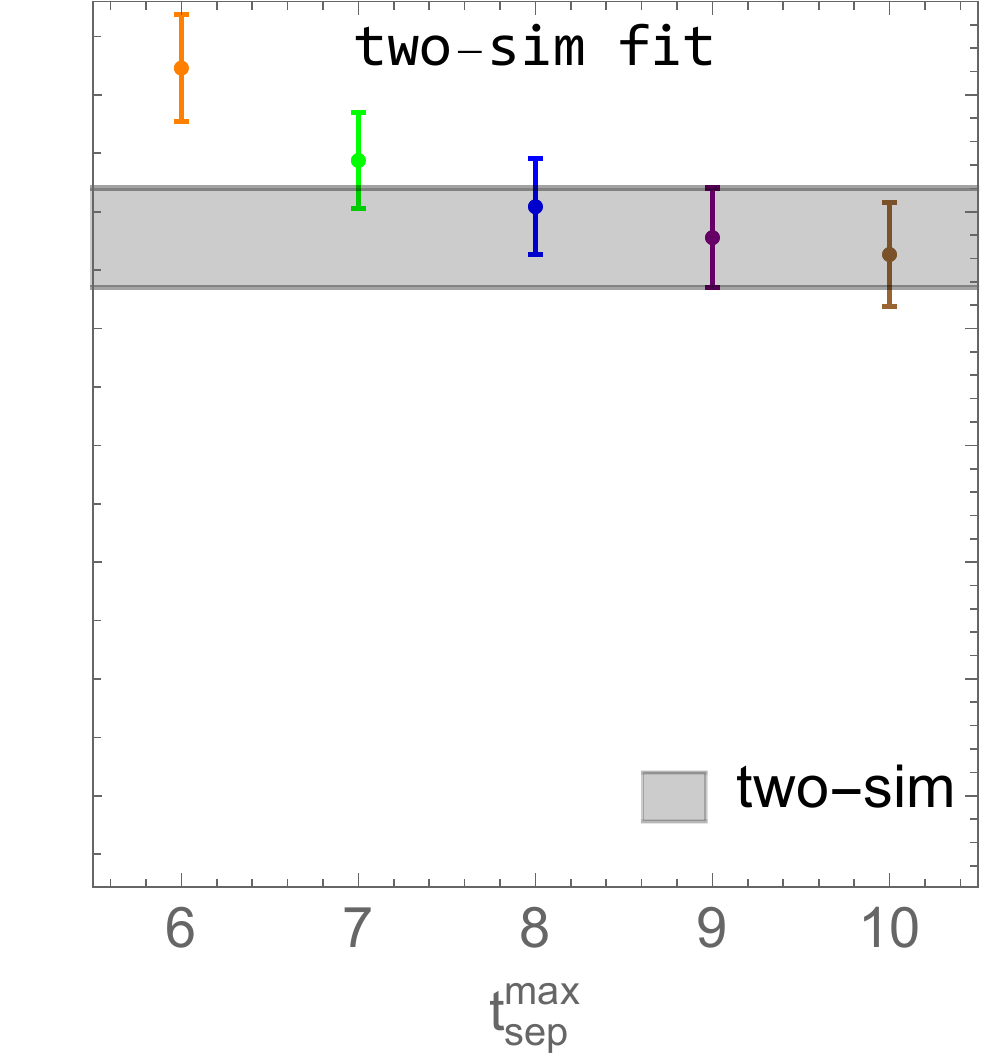}
\caption{Example ratio plots  (left column), 
and two-sim fits (last 2 columns) from the pion mass $a\approx 0.12$~fm, $M_\pi\approx 310$~MeV for $P_z=8\pi/L$, $z=4$ (upper row) and $P_z=4\pi/L$, $z=1$ (lower row).
The gray band shown in all plots corresponds to the extracted ground-state matrix element from the two-sim fit using $t_\text{sep}\in[5,9]$.
From left to right, the columns represent:
the ratio of the three-point to two-point correlators with the reconstructed fit bands from the two-sim fit using $t_\text{sep}\in [5,9]$, shown as functions of $t-t_\text{sep}/2$,
the two-sim fit results using $t_\text{sep}\in[t_\text{sep}^\text{min},9]$ as functions of $t_\text{sep}^\text{min}$, and
the two-sim fit results using $t_\text{sep}\in[5,t_\text{sep}^\text{max}]$
as functions of $t_\text{sep}^\text{max}$.
}
\label{fig:a12m310_Ratio_figure}
\end{figure*}

\begin{figure*}[!tbp]
  \centering
\includegraphics[width=0.35\textwidth]{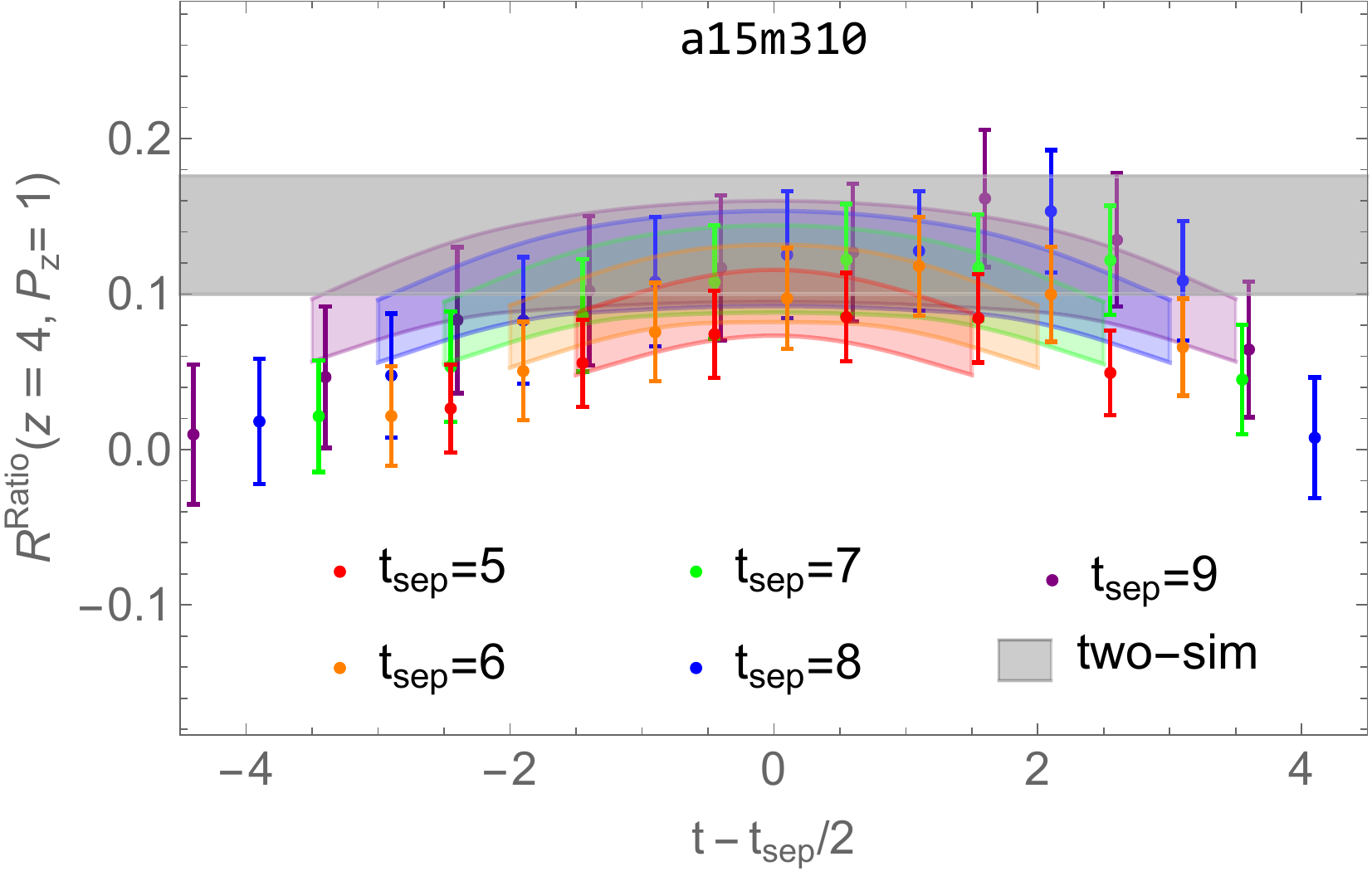}
\includegraphics[width=0.2\textwidth]{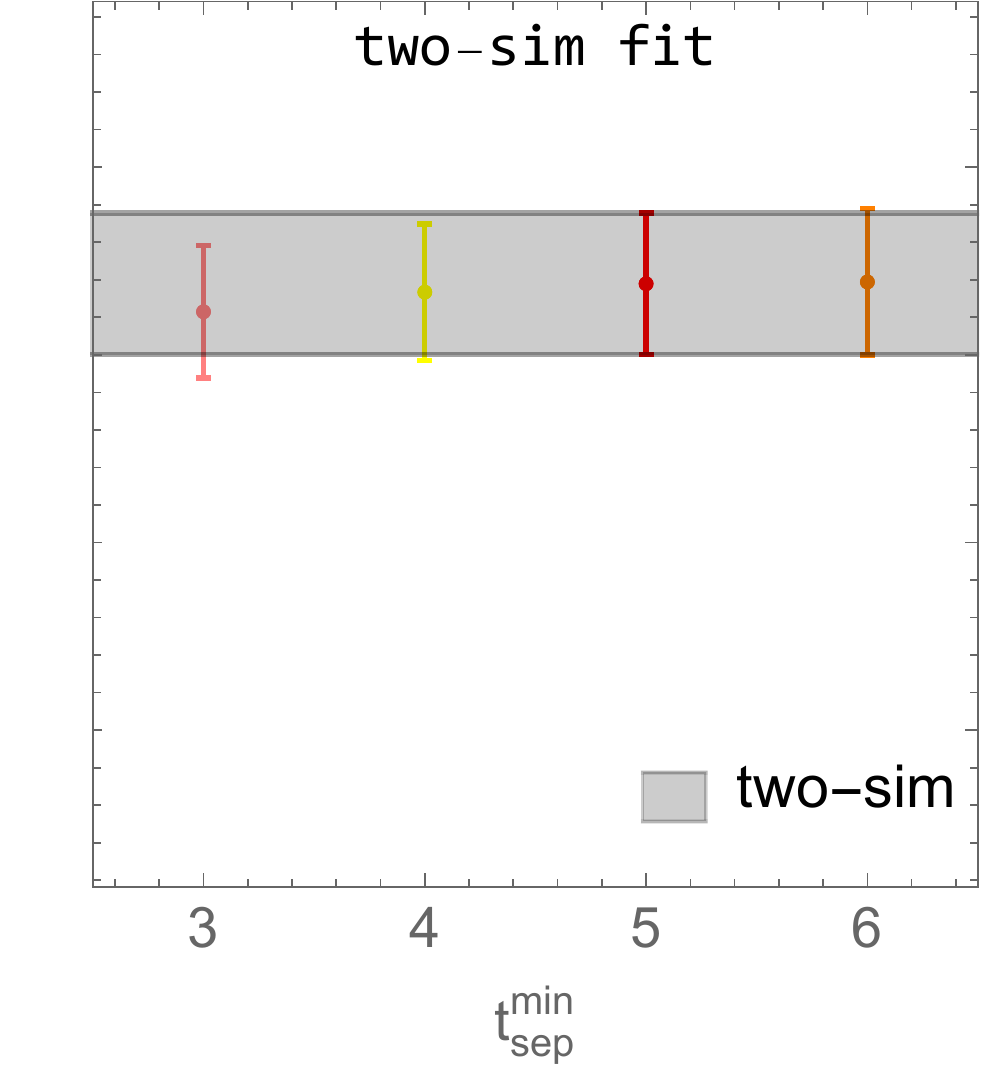}
\includegraphics[width=0.2\textwidth]{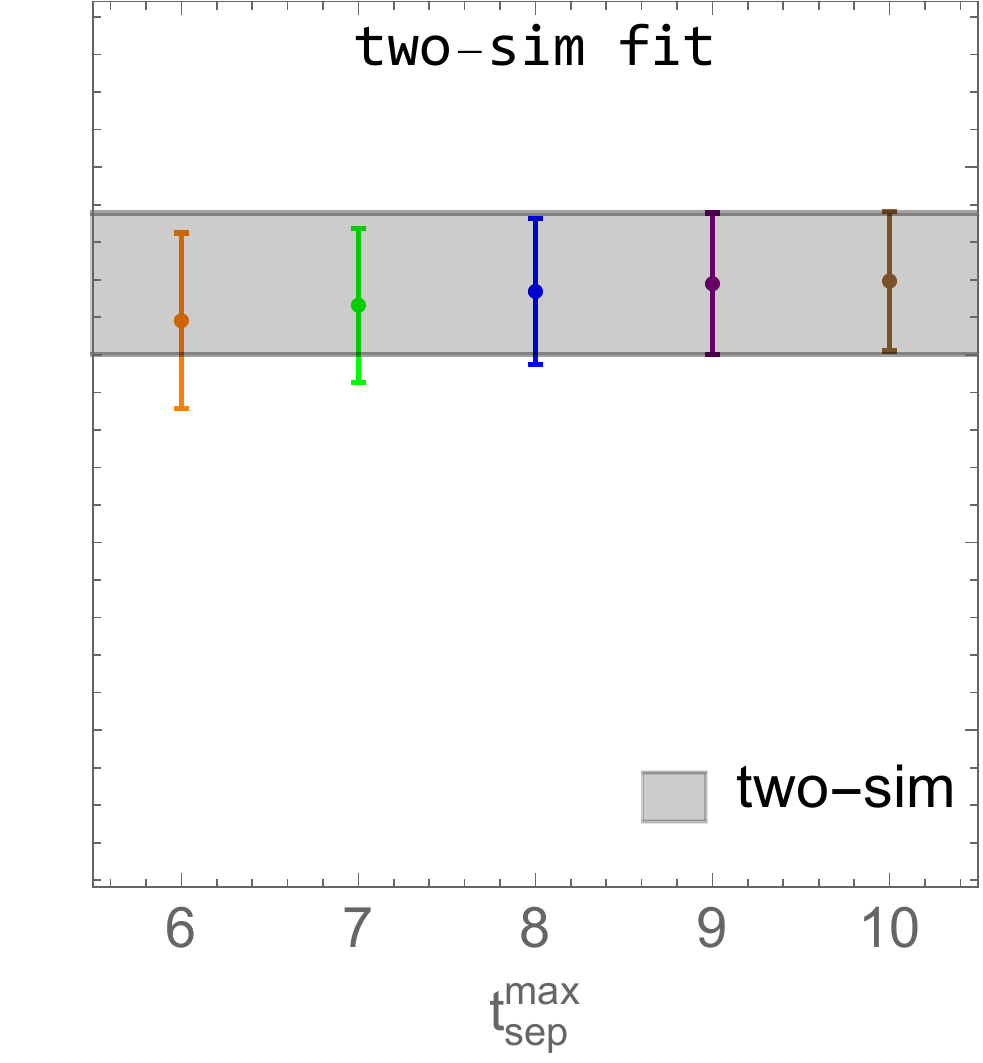}
\includegraphics[width=0.35\textwidth]{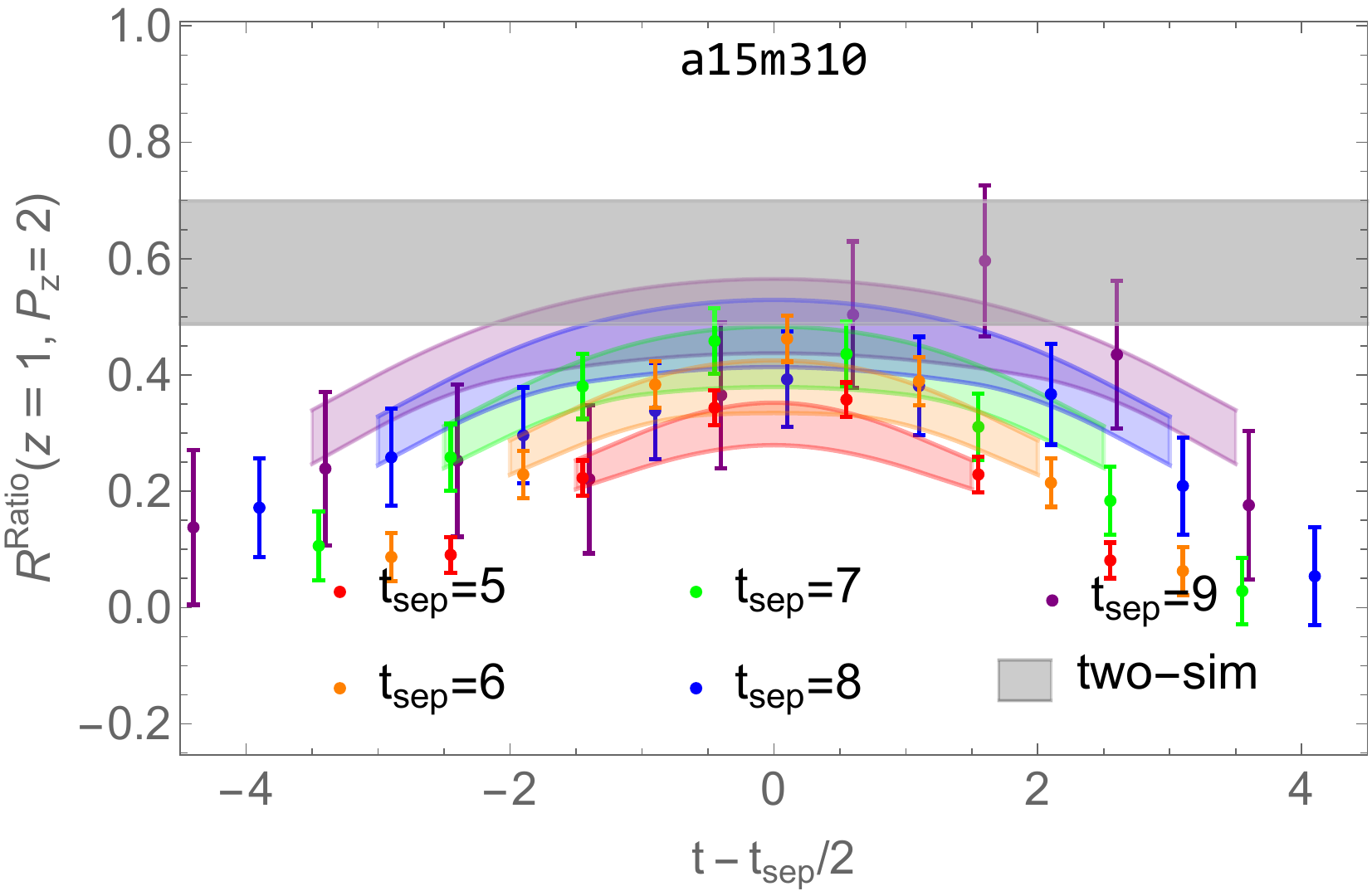}
\includegraphics[width=0.2\textwidth]{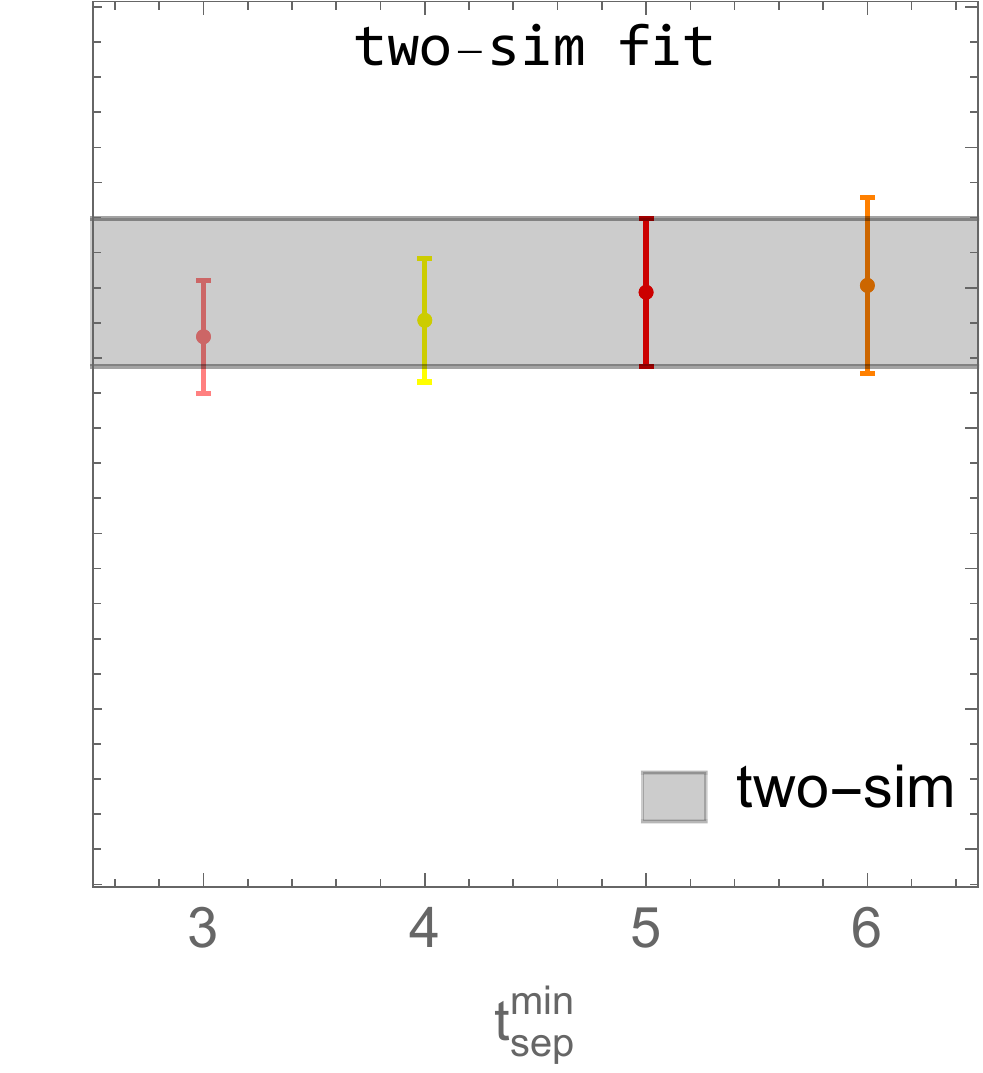}
\includegraphics[width=0.2\textwidth]{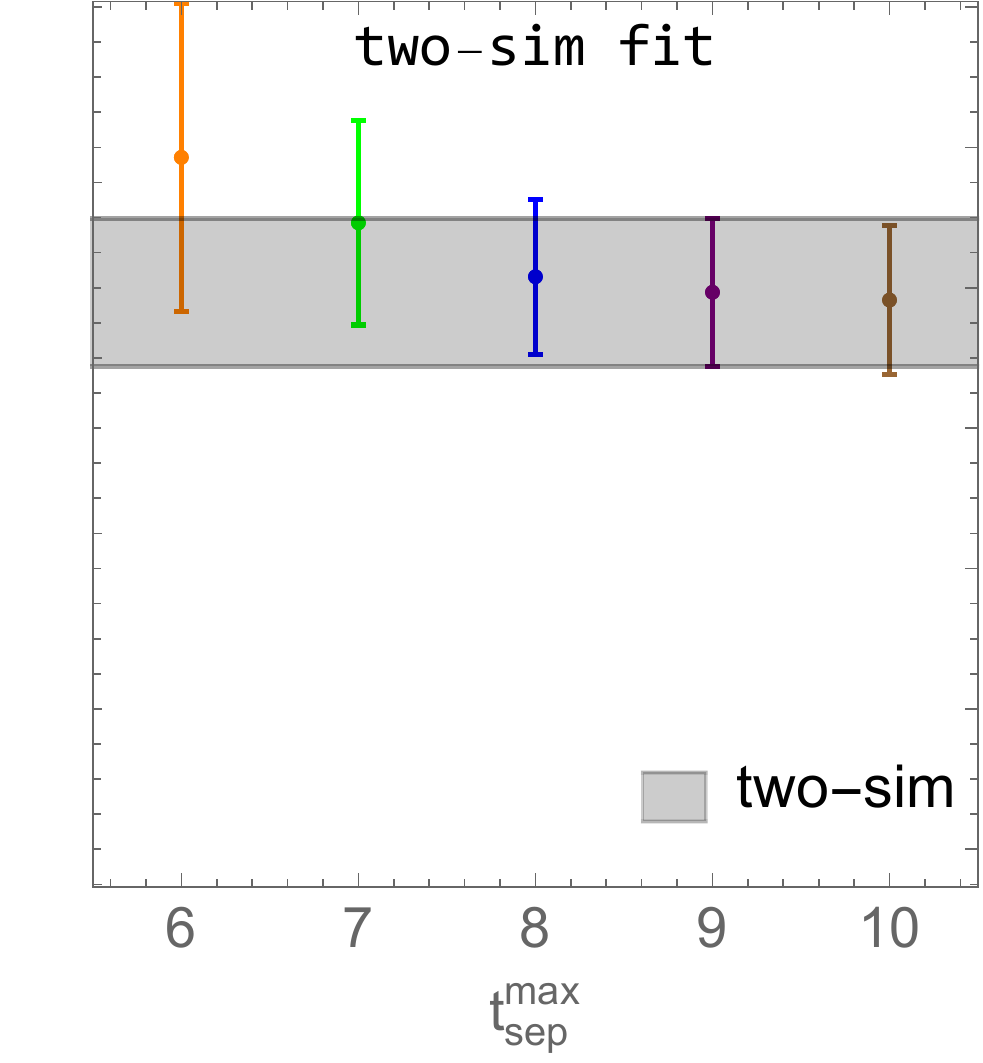}
\caption{Example ratio plots (left column),
and two-sim fits (last 2 columns) from the pion mass $a\approx 0.15$~fm, $M_\pi\approx 310$~MeV for $P_z=1\times 2\pi/L$, $z=1$ (upper row) and $P_z=4\times 2\pi/L$, $z=4$ (lower row).
The gray band shown in all plots corresponds to the extracted ground-state matrix element from the two-sim fit using $t_\text{sep}\in[5,9]$.
From left to right, the columns represent:
the ratio of the three-point to two-point correlators with the reconstructed fit bands from the two-sim fit using $t_\text{sep}\in [5,9]$, shown as functions of $t-t_\text{sep}/2$, 
the two-sim fit results using $t_\text{sep}\in[t_\text{sep}^\text{min},9]$ as functions of $t_\text{sep}^\text{min}$, and
the two-sim fit results using $t_\text{sep}\in[5,t_\text{sep}^\text{max}]$
as functions of $t_\text{sep}^\text{max}$.
}
\label{fig:a15m310_Ratio_figure}
\end{figure*}

The ground-state matrix element was obtained from a two-state fit to the three-point correlators or from a two-state simultaneous (two-sim) fit
across multiple time separations.
In order to corroborate that the fitted matrix elements were extracted correctly, the ratio
\begin{equation}\label{eq:Ratio}
R^\text{ratio}(z,P_z;t_\text{sep},t) = \frac{C^\text{3pt}(z,P_z;t_\text{sep},t)}{C^\text{2pt}(P_z;t_\text{sep})};
\end{equation}
of the three-point to the two-point correlator was computed.
The first column of Figs.~\ref{fig:a12m310_Ratio_figure} and \ref{fig:a15m310_Ratio_figure} show the kaon $R^\text{ratio}$ for the ensembles a12m310 and a15m310 for selected momenta $P_z$ and Wilson-line length $z$.
If there were no excited states, Eq.~\ref{eq:Ratio} would yield the ground-state matrix element.
It can be seen that $R^\text{ratio}$ increases as the source-sink separation increases. 
In some cases, $R^\text{ratio}$ starts to converge at large separation, which shows that the contribution from excited states diminishes.
The gray bands depict the ground-state matrix elements obtained from a ``two-sim'' fit to the three-point correlators at $t_\text{sep}\in [5,9]$.
In the second and third columns, one can see the variation of the fitted ground-state matrix elements for different choices of $t_\text{sep}^\text{min}$ and $t_\text{sep}^\text{max}$.
In the examples shown in Fig.~\ref{fig:a15m310_Ratio_figure}, our ground-state fitted matrix elements are stable for the a15m310 ensemble regardless of the choices we made for $t_\text{sep}^{\text{min},\text{max}}$.
However, for the example from the a12m310 ensemble, we found that the ground-state matrix elements are sensitive to the selection of $t_\text{sep}$.
Our choices of $t_\text{sep}\in [5,9]$ for the analysis do provide a reliable ground-state extraction.

\section{Kaon Gluon PDFs from the Pseudo-PDF Method}\label{sec:results}

The Ioffe-time pseudo-distribution (pITD)~\cite{Radyushkin:2017cyf,Orginos:2017kos} is given by
\begin{equation}
\mathcal{M}(\nu,z^2) = \langle 0 (P_z)|{\cal O}(z)|0 (P_z)\rangle,
\label{eq:ME_unpol}
\end{equation}
The reduced pITD (RpITD)~\cite{Orginos:2017kos,Zhang:2018diq,Li:2018tpe} was constructed to remove the ultraviolet divergences in the pITD by taking the ratio of the pITD to its corresponding $z$-dependent matrix element with $P_z=0$, then normalizing the ratio by the matrix element at $z^2=0$,
\begin{equation}
\mathscr{M}(\nu,z^2)=\frac{\mathcal{M}(zP_z,z^2)/\mathcal{M}(0\cdot P_z,0)}{\mathcal{M}(z\cdot 0,z^2)/\mathcal{M}(0\cdot 0,0)}.
\label{eq:RITD}
\end{equation}
The renormalization of ${\cal O}(z)$ and kinematic factors are cancelled in the RpITDs.
By construction, the RpITD double ratios employed here are normalized to one at $z=0$.

\begin{figure}[!tbp]
\centering
\includegraphics[width=0.45\textwidth]{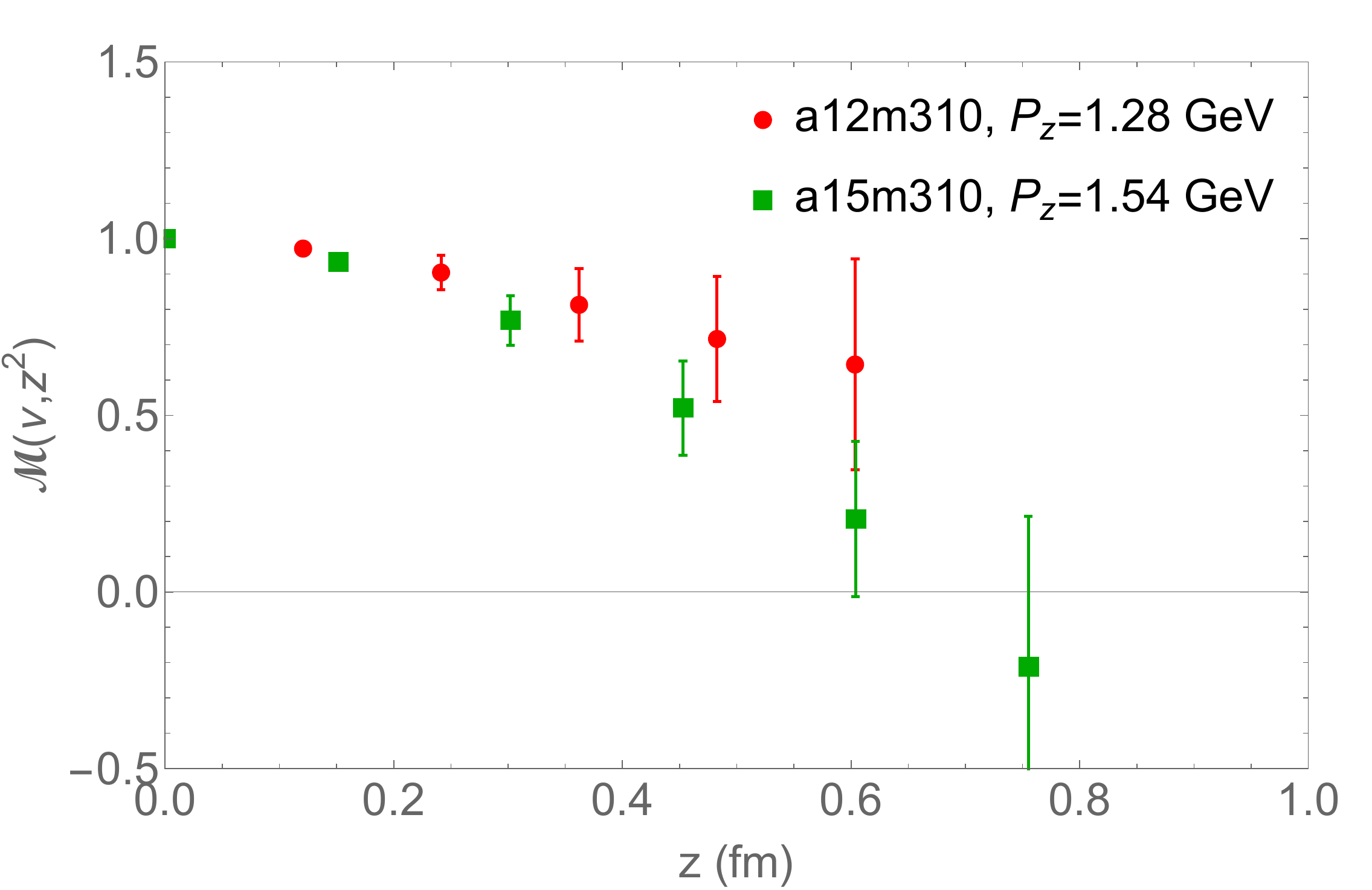}
\caption{\label{fig:ensemble_momentum_comparison}
Kaon RpITDs for the a12m310 and a15m310 ensembles at boost momentum $P_z \approx 1.3$~GeV as a function of the Wilson-line displacement $z$.
}
\end{figure}

Figure~\ref{fig:ensemble_momentum_comparison} shows the lattice-spacing dependence of the kaon gluon RpITD.
The RpITDs are plotted as a function of Wilson link length $z$ at approximately the same momentum $P_z$ for ensembles a12m310 and a15m310, which have lattice spacings around 0.12 and 0.15~fm.
We observe that the RpITD from the a12m310 ensemble has higher values for each  Wilson-link length and has a slower rate in decrease of its values.
Most points are within one sigma of their corresponding points on the other ensemble, with slightly larger central values for most $z$.
As the Wilson-link length increases, we see the difference increases.

\begin{figure*}[!tbp]
\centering
\includegraphics[width=0.32\textwidth]{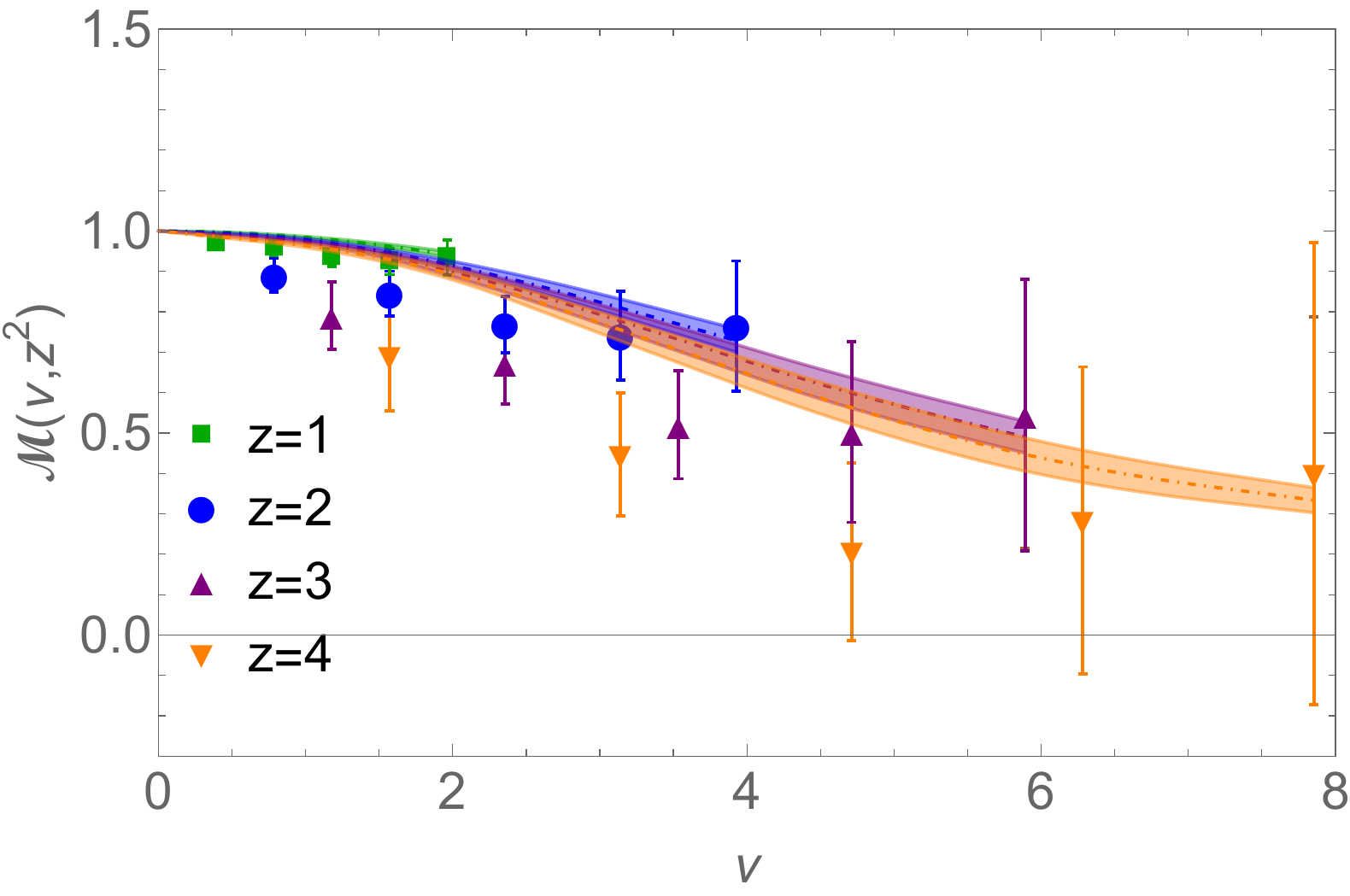}
\includegraphics[width=0.32\textwidth]{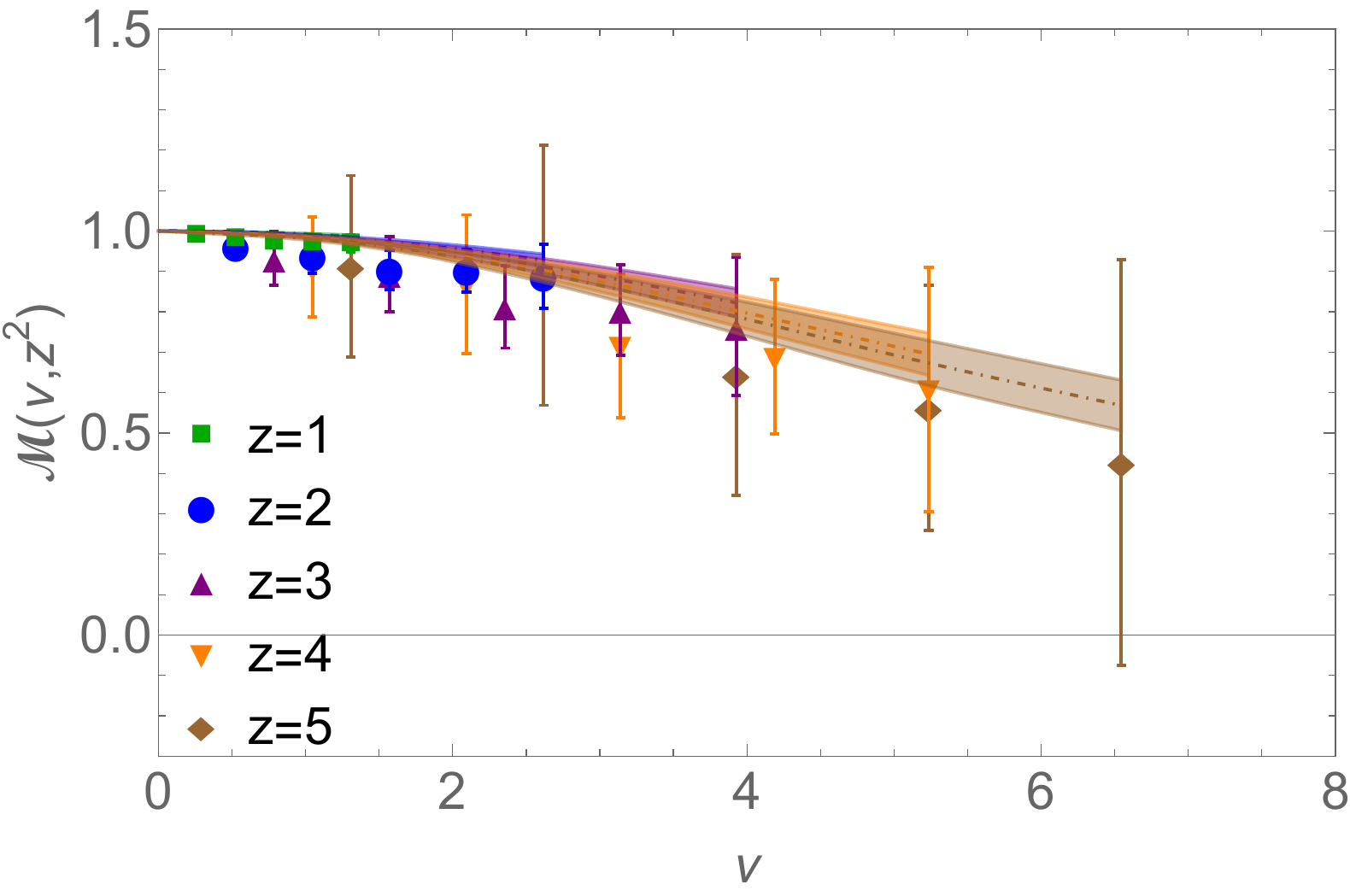}
\includegraphics[width=0.32\textwidth]{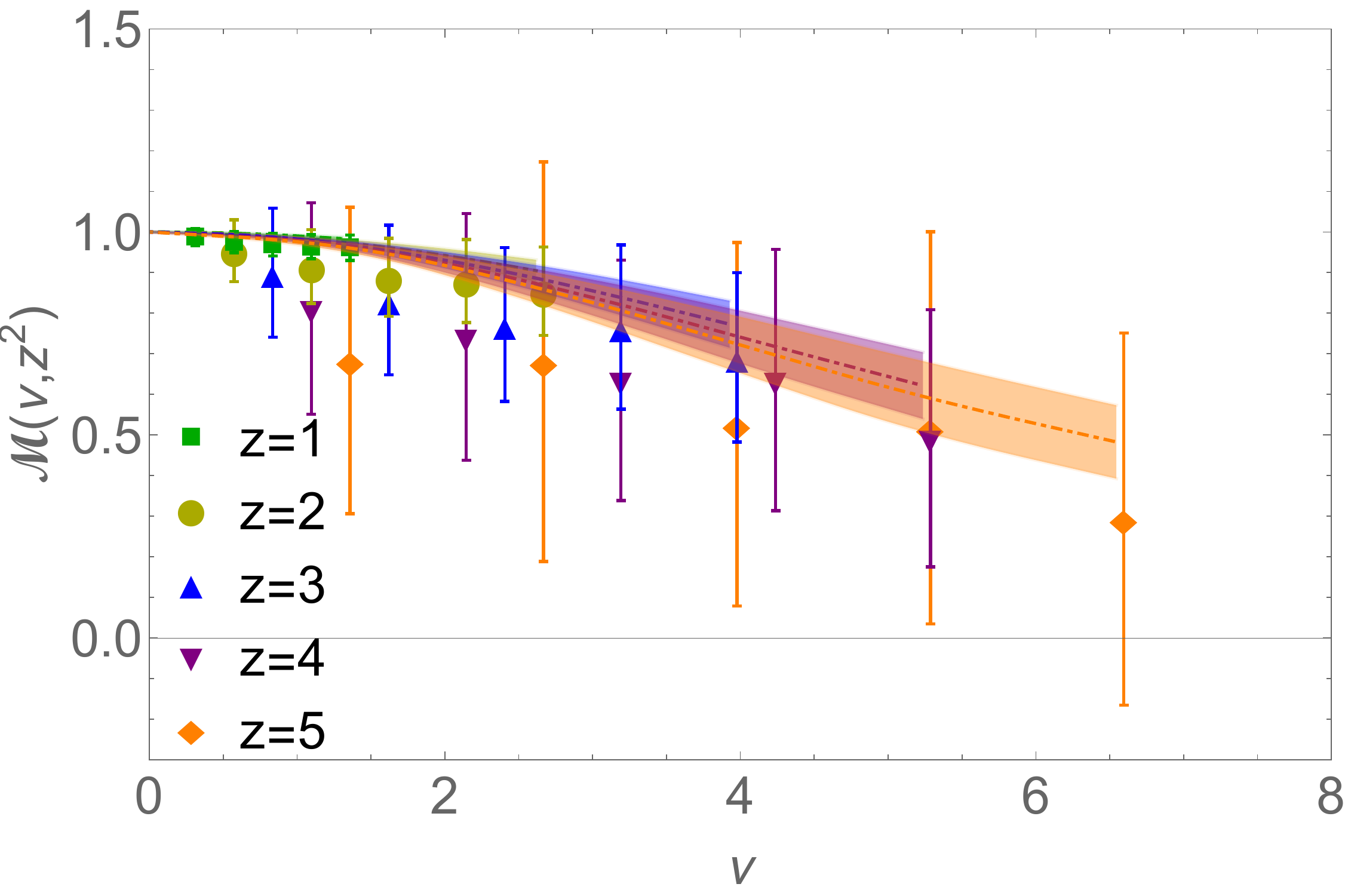}
\caption{\label{fig:RITD_fitbands}
Kaon RpITD for the a15m310 (left), a12m310 (middle) ensembles, and pion RpITD for a12m310 (right).
The bands are the gluon PDF fits to each $z$ by minimizing the $\chi^2$ defined in Eq.~\ref{eq:chi2-RITD-fit} to obtain the gluon PDFs.}
\end{figure*}

Using the matrix elements obtained from Eq.~\ref{eq:3ptC} and Eq.~\ref{eq:RITD}, we compute the RpITD for both ensembles at the various Wilson link lengths $z$ and meson momenta $P_z$.
Figure~\ref{fig:RITD_fitbands} shows these results compared to the RpITD previously calculated for the pion on the a12m310 ensemble in Ref.~\cite{Fan:2021bcr}.
We see that the kaon RpITDs for both ensembles, shown in the left and middle plots, are congruent.
Additionally, the RpITD for the kaon and pion, shown in the rightmost plot, from a12m310 are very similar, in accordance with the results from Ref.~\cite{Roberts:2021nhw}, where the ratio of the kaon PDF to the pion PDF was calculated and found to be around 1.

We can then extract the gluon PDF distribution through the pseudo-PDF matching condition~\cite{Balitsky:2019krf} that connects the RpITD $\mathscr{M}$ to the lightcone gluon PDF $g(x,\mu^2)$ through
\begin{align}
\mathscr{M}(\nu,z^2)&=\int_0^1 dx \frac{xg(x,\mu^2)}{\langle x \rangle_g}R_{gg}(x\nu,z^2\mu^2),
\label{eq:matching-gg}
\end{align}
where $\mu$ is the renormalization scale in the $\overline{\text{MS}}$ scheme
and $\langle x \rangle_g=\int_0^1 dx \, x g(x,\mu^2)$ is the gluon momentum fraction of the kaon.
$R_{gg}$ is the gluon-in-gluon matching kernel
\begin{widetext}
\begin{equation}
R_{gg}(y,z^2,\mu^2)
=\cos y -\frac{\alpha_s(\mu)}{2\pi}N_c
\left\{ \left[\ln\left(z^2\mu^2\frac{e^{2\gamma_E+1}}{4}\right)+2\right]R_B(y)+R_L(y)+R_C(y)\right\}, \nonumber
\label{eq:Rgg}
\end{equation}
\end{widetext}
where $\alpha_s$ is the strong coupling at scale $\mu$,
$N_c=3$ is the number of colors,
and $\gamma_E=0.5772$ is the Euler-Mascheroni constant.
For the term $R_{gg}(y,z^2,\mu^2)$, $z$ was chosen to be $2e^{-\gamma_E-1/2}/\mu$ so that the logarithmic term vanishes, which suppresses the residuals containing higher order logarithmic terms, following the previous publication regarding the one-loop evolution of the pseudo-PDF~\cite{Radyushkin:2018cvn}.
Equation~\ref{eq:matching-gg} and the terms $R_B(y)$, $R_L(y)$, $R_C(y)$ are defined in Eqs.~7.21--23 and in the paragraph below Eq.~7.23 in Ref.~\cite{Balitsky:2019krf}.

Note that the lattice-calculated RpITDs are also connected to the singlet quark-PDF $q_s$ of the kaon via the quark-gluon matching kernel $R_{gq}$ with an additional $\frac{P_z}{P_0}\int_0^1 dx \frac{xq_S(x,\mu^2)}{\langle x \rangle_g}R_{gq}(x\nu,z^2\mu^2)$ term added to Eq.~\ref{eq:matching-gg}.
In the past pion gluon PDF study~\cite{Fan:2021bcr}, it was found that the quark PDF contribution is small;
therefore in this work, we will neglect the kaon quark PDF.
One can obtain the gluon PDF $g(x,\mu^2)$ by fitting the RpITD through the matching condition in Eq.~\ref{eq:matching-gg};
a similar procedure has also been used by HadStruc Collaboration~\cite{Joo:2019bzr,Joo:2020spy,HadStruc:2021wmh}.

To obtain the gluon PDF $g(x,\mu^2)$ on the right-hand side of Eq.~\ref{eq:matching-gg}, we adopt the phenomenologically motivated form
\begin{align}
f_g(x,\mu) = \frac{xg(x, \mu)}{\langle x \rangle_g(\mu)} = \frac{x^A(1-x)^C}{B(A+1,C+1)},
\label{functional}
\end{align}
for $x\in[0,1]$ and zero elsewhere.
The beta function $B(A+1,C+1)=\int_0^1 dx\, x^A(1-x)^C$ is used to normalize the area to unity.
Such a form is also used in global fits to obtain the pion gluon PDF by JAM~\cite{Barry:2018ort,Cao:2021aci}.
We fit the lattice RpITDs $\mathscr{M}^\text{lat} (\nu,z^2,a,M_\pi)$ obtained in Eq.~\ref{eq:RITD} to the parametrization form $\mathscr{M}^\text{fit}(\nu,\mu,z^2,a,M_\pi)$ in Eq.~\ref{eq:matching-gg} by minimizing the $\chi^2$ function,
\begin{equation}
\label{eq:chi2-RITD-fit}
    \begin{split}
    & \chi^2(\mu,a,M_\pi)=\\
    & \sum _{\nu,z} \frac{(\mathscr{M}^{\rm fit}(\nu,\mu,z^2,a,M_\pi)-\mathscr{M}^{\rm lat}(\nu,z^2,a,M_\pi))^2}{\sigma^2_{\mathscr{M}}(\nu,z^2,a,M_\pi)}.
    \end{split}
\end{equation}
The reconstructed fit bands of the kaon RpITDs for the a15m310 and a12m310 ensembles, and pion RpITDs for a12m310 at each $z^2$, compared with the lattice calculation points, are shown in Fig.~\ref{fig:RITD_fitbands} from left to right.
We see almost no $z^2$-dependence (labeled in different colors) in the reconstructed bands in both a12m310 ensembles, but slightly more dependence in the a15m310 case.
The RpITD data points fluctuate more in the a15m310 kaon data and as a result the fit quality is not as good as in the a12m310 case. 
We found the a12m310 fit to both the pion and kaon gluon PDF to have very stable quality with $\chi^2/\text{dof}$ around 1 with consistent output of $f_g(x,\mu)$, regardless of the choice of the maximum value of the Wilson-line displacement $z$.
However, for the a15m310 ensemble, $\chi^2/\text{dof}$ can go as large as 9(3).
By reducing the fitted $z$ to smaller number $z=1$, $\chi^2/\text{dof}$ can be reduced to 5(3).
We suspect that higher-twist effects are enhanced at this coarse lattice spacing such that the fit fails to accurately describe the lattice data.
Possible future work including NNLO matching may help to improve the fit on this ensemble.

\begin{figure*}[!tbp]
  \centering
\includegraphics[width=0.45\textwidth]{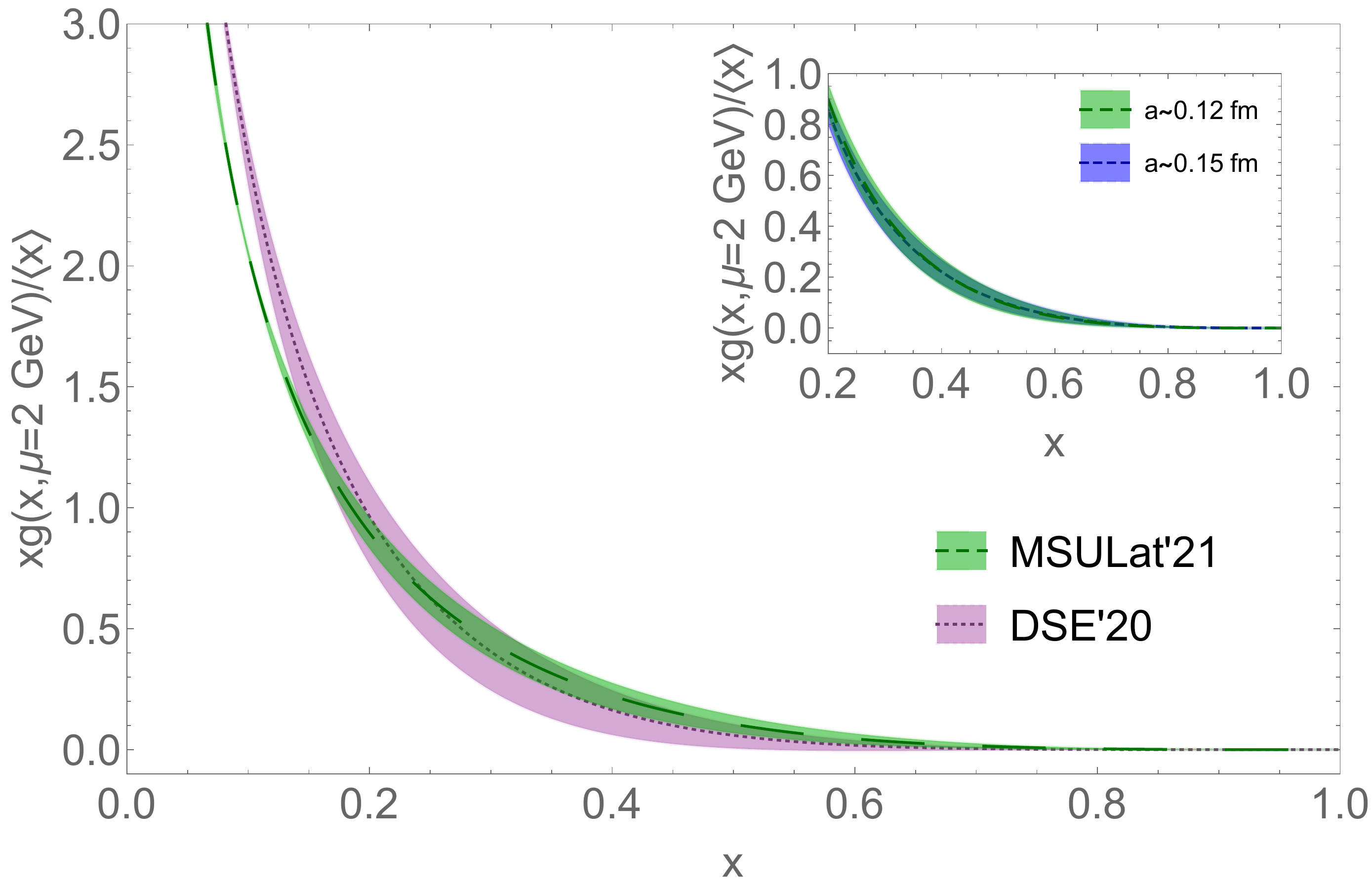}
\includegraphics[width=0.45\textwidth]{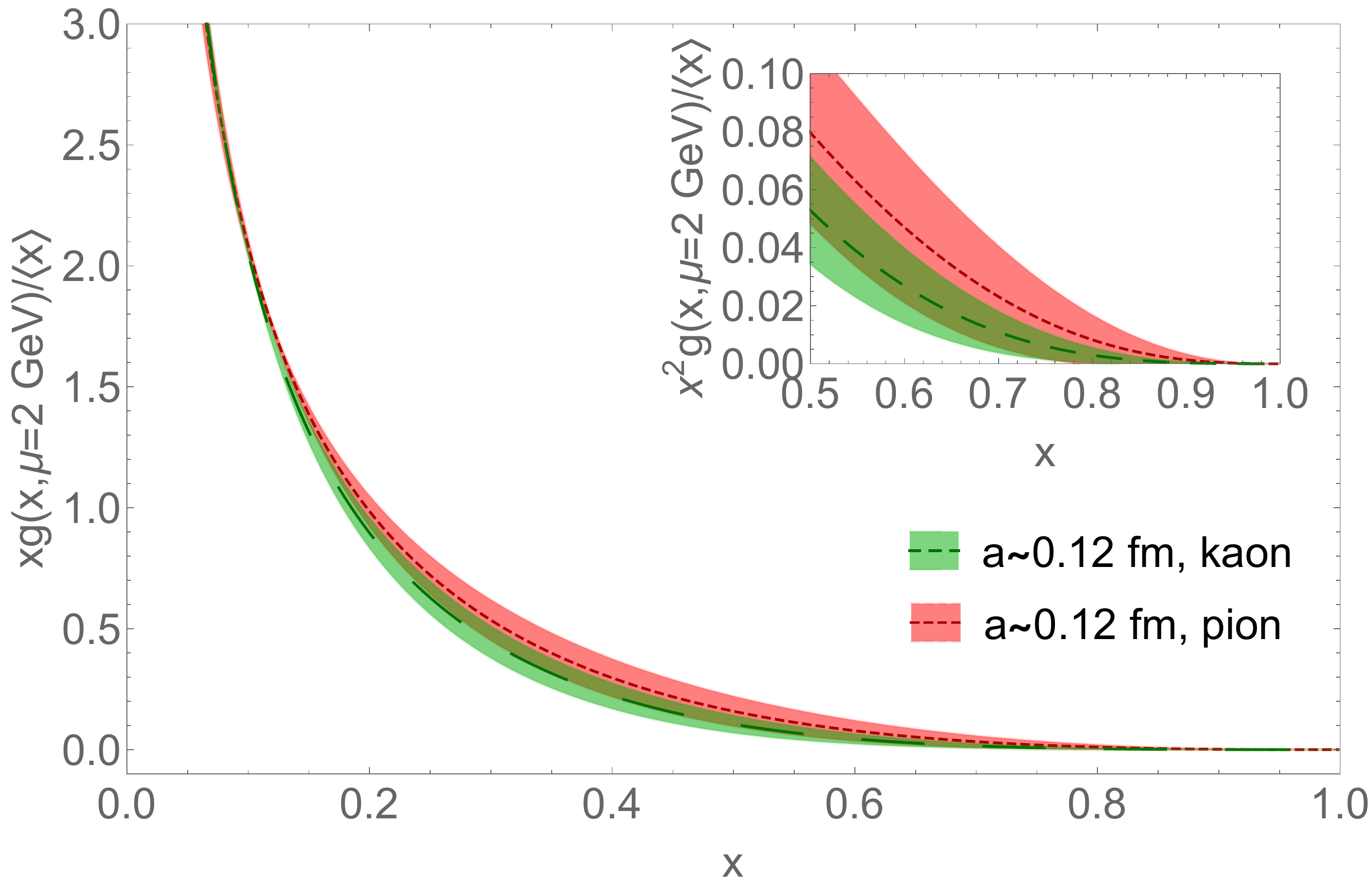}
\caption{\label{fig:DSE_comparison}
Left plot: The kaon gluon PDF $xg(x, \mu)/\langle x \rangle_g$ as a function of $x$ obtained from the fit to the lattice data on ensembles with lattice spacing $a\approx\{0.12,0.15\}$~fm (inset plot), pion masses $M_\pi\approx310$~MeV at $a\approx 0.12$~fm, compared with the kaon gluon PDF from DSE'20 at $\mu=2$~GeV in the $\overline{\text{MS}}$ scheme. 
Right plot: Comparison of pion and kaon gluon PDF $xg(x, \mu)/\langle x \rangle_g$ as a function of $x$ with lattice spacing $a\approx0.12$~fm (outer) and 0.15~fm (inset), pion masses $M_\pi\approx310$~MeV, at $\mu=2$~GeV in the $\overline{\text{MS}}$ scheme.
}
\label{fig:xgx}
\end{figure*}

Our results on the kaon and pion gluon PDFs $xg(x, \mu)/\langle x \rangle_g$ as a function of $x$ from fitting the lattice data are shown in Fig.~\ref{fig:xgx}.
In the inset of the left-hand plot, we show the kaon $xg(x, \mu)/\langle x \rangle_g$ at $\mu=2$~GeV in the $\overline{\text{MS}}$ scheme with $M_\pi\approx 310$~MeV from two lattice spacings $a\approx\{0.12,0.15\}$~fm.
Note that we use $z=1$ for the coarser lattice spacing due to the fit quality becoming significantly worse as $z$ increases, whereas we are able to get a good fit for the finer-lattice RpITD data for $z$ up to 5.
With these choices of data input, our best determination of the kaon $xg(x, \mu)/\langle x \rangle_g$ are consistent within the statistical errors.  
We also compare our finer-lattice result with the same quantity obtained from the Dyson-Schwinger equation (DSE) approach ~\cite{Cui:2020tdf}, depicted as the purple band on the left-hand side.
Reference~\cite{Cui:2020tdf} provides recent DSE calculations of the pion gluon PDF and the ratio of kaon to pion gluon PDFs as function of $x$.
Although the kaon gluon PDF is not directly provided, it can be obtained from the product of the pion gluon PDF and the kaon-to-pion ratio.
Accordingly, we can make a direct comparison between our lattice calculations and the kaon gluon PDF from DSE calculations.
On the other hand, the DSE'20 error shown is likely overestimated due to lack of correlations in the published parameters for their kaon and pion PDFs.
Our kaon gluon PDF results from the smallest lattice spacing, $a\approx 0.12$~fm, is consistent with the DSE results~\cite{Cui:2020tdf} within one-sigma error for $x>0.15$.
On the right-hand side of Fig.~\ref{fig:xgx}, we compare the obtained kaon result at smaller lattice spacing with the pion results obtained from the same ensembles;
we note the kaon gluon PDF is slightly smaller than the one obtained for the pion, similar to its corresponding quark valence PDF and DSE~\cite{Cui:2020tdf} results. 
We predict the kaon $\langle x^2 \rangle_g/\langle x \rangle_g$ and  $\langle x^3 \rangle_g/\langle x \rangle_g$ to be 0.0779(94) and 0.0187(42), while in good agreement with corresponding results from DSE~\cite{Cui:2020tdf}: 0.075 and 0.015. 
On the pion $\langle x^2 \rangle_g/\langle x \rangle_g$, our 310-MeV results gives 
 .092(15) and 0.0250(75), 
while results from DSE~\cite{Cui:2020tdf},  JAM~\cite{Barry:2018ort,Cao:2021aci} and xFitter~\cite{Novikov:2020snp}
are 
0.076, 0.103, 0.158  
and 0.015, 0.024, 0.048,
respectively. 
Future study including finer lattice spacing and lighter pion mass will be important to refine this calculation and provide better predictions on this poorly known meson quantity.

\section{Summary}\label{sec:summary}
In this work, we made the first lattice study of kaon gluon parton distribution, using the pseudo-PDF approach.
We used clover fermions as valence action and 310-MeV 2+1+1 HISQ configurations generated by the MILC collaboration at two lattice spacings, 0.15 and 0.12~fm.
We used momentum smearing and high-statistics measurements, up to O(324,000), to reach the kaon boost momentum around 2~GeV.
We carefully studied the excited-state contributions to the matrix elements using a two-state fitting strategy and made sure that our ground-state matrix elements were stably obtained.
We then calculated the reduced pseudo-ITD using the obtained fitted ground-state matrix elements and extracted the gluon parton distribution.
We found that the kaon gluon PDF at the finer lattice spacing is consistent with the DSE  result~\cite{Cui:2020tdf} within statistical uncertainties, except in the small-$x$ region, which our $z P_z$ is too small to constrain.
When comparing with the pion PDF result, we found the kaon PDF to be slightly smaller in central value for most of the $x > 0.2$ region.
We found that the kaon gluon PDFs show potential discretization error at the coarse lattice spacing of 0.15~fm;
future study using an additional lattice spacing of 0.09~fm would give us a better estimate of the systematics uncertainty in the 0.12~fm results.
We suspect the quark-gluon mixing is smaller than our statistical error based on the prior pion gluon calculation.
Other systematics from sources such as higher-twist contributions, finite-volume effects, unphysical pion-mass effects are not included in this pioneer study.
Future studies should aim at improving these systematics and provide a better determination of the kaon gluon PDFs for the upcoming experimental efforts.

\section*{Acknowledgments}
We thank MILC Collaboration for sharing the lattices used to perform this study.
The LQCD calculations were performed using the Chroma software suite~\cite{Edwards:2004sx}.
This research used resources of the National Energy Research Scientific Computing Center, a DOE Office of Science User Facility supported by the Office of Science of the U.S. Department of Energy under Contract No. DE-AC02-05CH11231 through ERCAP;
facilities of the USQCD Collaboration, which are funded by the Office of Science of the U.S. Department of Energy,
and supported in part by Michigan State University through computational resources provided by the Institute for Cyber-Enabled Research (iCER).
The work of AS, ZF and HL are partially supported by the US National Science Foundation under grant PHY~1653405 ``CAREER: Constraining Parton Distribution Functions for New-Physics Searches''.


\end{document}